\documentclass[%
 reprint,
 amsmath,amssymb,
 aps,
]{revtex4-1}

\usepackage{graphicx}
\usepackage{dcolumn}
\usepackage{bm}
\usepackage[colorlinks=true]{hyperref}
\usepackage{siunitx}
\usepackage{subfigure}
\usepackage[makeroom]{cancel}

\usepackage{xcolor}
\usepackage{soul}
\usepackage{systeme}
\newcommand{\erfc}{\text{erfc}}
\newcommand{\erfi}{\text{erfi}}

\newcommand{\amended}[1]{\color{black}#1\color{black}}

\begin{document}
\title{
Electrolytes structure near electrodes with molecular size roughness
}

\author{
Timur Aslyamov
\\
\texttt{t.aslyamov@skoltech.ru}
\\
Center for Design, Manufacturing and Materials, Skolkovo Institute of Science and Technology, Bolshoy Boulevard 30, bld. 1, Moscow, Russia 121205
\\
Konstantin Sinkov\\
\texttt{sinkovk@gmail.com}\\
Schlumberger Moscow Research, Leningradskoe shosse 16A/3, Moscow, Russia 125171\\
Iskander Akhatov\\
Center for Design, Manufacturing and Materials,
Skolkovo Institute of Science and Technology,
Bolshoy Boulevard 30, bld. 1, Moscow, Russia 121205
}


\date{\today}

\begin{abstract}
Understanding the electrodes' surface morphology influence on the ions' distribution is essential for designing the supercapacitors with enhanced energy density characteristics. We develop a model for the structure of electrolytes near the rough surface of electrodes. 
The model describes an effective electrostatic field's increase and associated intensification of ions' spatial separation at the electrode-electrolyte interface.
These adsorption-induced local electric and structure properties result in notably increased values and sharpened
form of the DC dependence on the applied potential. Such capacitance behavior is observed in many published simulations, and its description is beyond the capabilities of the established flat-electrodes theories. The proposed approach could extend the quantitatively verified models providing a new instrument of the electrodes surface-parameters optimization for specific electrolytes.  
\end{abstract}

\maketitle
The supercapacitors are one of the most prospective modern energy sources due to the outstanding charging/discharging times, extremely long life-cycle, and ecology-friendly process of the charge storage \cite{simon2020perspectives}.
\amended{
Since supercapacitors store the electrical energy using the ions adsorption, the porous carbon materials with high specific surface area (SSA) are used as popular electrodes \cite{gamby2001studies, chmiola2006effect, harmas2020hydrothermal}. 
Such materials exhibit wide pores size distribution dividing a large pores volume between micro- and meso-scales \cite{beguin2014carbons}.
}
However, the vital experimental discovery of the capacitance increase inside sub-nanoporous electrodes \cite{chmiola2006anomalous} shows that not all pores contribute effectively to the capacity. 
Further experimental \cite{largeot2008relation} and theoretical \cite{jiang2011oscillation} studies demonstrated that the capacitance of the nanoporous electrodes is an oscillating function of the size of the pores with the largest value corresponding to pore's size comparable to ions diameter.
\amended{
This anomalous enhancement of the capacity demands the dense packing of the identically charged ions in spite of the electrostatic repulsion. It was explained and simulated in works \cite{kondrat2010superionic, kondrat2011superionic} accounting for the polarisation of the conductive pores boundaries which screens the electrostatic ion-ion interaction.
These results show the importance of the molecular-size influence on the electrode-electrolyte interface storing the energy in the supercapacitors based technologies.
Thus, the molecular scale surface roughness which is inherent for the carbons mesopores \cite{sheehan2016layering} could crucially influence the capacity properties.
}

\amended{
The investigation of the morphology's influence on the electrical properties has a long history starting with the experiments with solid metal electrodes (see \cite{vorotyntsev1986modern} for review).
The first models have be developed by Daikhin, Kornyshev and Urbakh for the description of the diluted electrolytes near the electrodes with a weak roughness. The authors used the perturbation theory for both the linear \cite{daikhin1996double} and nonlinear \cite{daikhin1998nonlinear} Poisson-Boltzmann equations. The electrostatic fields perturbed by the roughness have been shown to result in the increase of the capacity in comparison with the flat electrodes. 
}
The systematic molecular dynamic (MD) studies \cite{vatamanu2011influence,xing2012nanopatterning,hu2013molecular,vatamanu2014influence, bedrov2015ionic, vatamanu2017charge} demonstrated the enhanced capacitance for the carbon electrodes with the molecular-scale patterns on the surface.
\amended{
Very recently the importance of the morphology has been clearly illustrated by the experimental comparison of two porous carbon samples with almost equal SSA, porous size distribution and composition, but the different surface roughness \cite{wei2020surface}. The measured capacitance of the material with higher roughness is more than 50\% higher than more smoother electrode. 
Thus, the adoption of the nanosize pores is not the only direction to enhance the capacitance, another option is using the electrodes with geometrically heterogeneous surfaces \cite{vatamanu2015non}.
}


The differential capacitance (DC) $C_d$ depending on the applied potential $U$ is traditionally used as the main characteristic to quantify the geometrical influence on the electrostatic properties in the experiments \cite{su2009double, lauw2012structure} and simulations mentioned above. 
In the case of flat electrodes, Kornyshev \cite{kornyshev2007double} predicted that DC as a function of potential exhibits camel- or bell-shapes in the dependence on the electrolyte packing density.
\amended{
The extension of this mean-field model accounting for inter-molecular interactions \cite{goodwin2017mean} provide more realistic DC's properties: smoother form and reduced maximum's values for the DC dependence on the potential. Also, similar mean-field density functional was used to estimate the number ratio of the bounded- (neutral clusters) and free-states of the particles in the ionic liquids \cite{feng2019free}.
}
In the case of rough electrodes the computer simulations indicate a more complicated DC behavior, for example, the MD study \cite{vatamanu2011influence} showed that the electrode surface's geometrical heterogeneity could alter the form of the ionic liquids DC from the camel-shape to bell-shape. Also, in work \cite{xing2012nanopatterning} the authors considered significantly rough surfaces which induce a sharper form of DC and the existence of a larger number of peaks. 
\amended{
Thus, to describe the rough surface influence on the DC the desired  theory should extend the surface description beyond flat/curved geometries \cite{kornyshev2007double, gupta2020ionic, chao2020effects, janssen2019curvature} and account for more realistic concentrate electrolyte properties than used in pioneering works \cite{daikhin1996double, daikhin1998nonlinear} 
}

In this work, we develop a mean-field model describing the electrolyte behavior near rough electrode surfaces.  
\amended{
Here we focus on the ionic liquids \cite{fedorov2014ionic} exhibiting the complex behaviour of DC in the dependence of the surface roughness \cite{vatamanu2017charge, vatamanu2015non}. 
In order to demonstrate the influence of the electrodes roughness on the DC properties more explicitly we consider simple mean field model \cite{kornyshev2007double} as a basis (the flat electrodes limit).
The mean-filed models \cite{kornyshev2007double, goodwin2017mean} describe the structure-less "crowding" state only. Description of the well-structured "overscreening" demands models beyond mean-field, for example, \cite{bazant2011double, de2020interfacial}.
However, the oscillating charge behaviour indicating the overscreening has been observed in the experiments with the atomically flat electrodes only \cite{goodwin2017mean}. 
Moreover, as it is known from theory \cite{neimark2009quenched} and experiment \cite{sheehan2016layering} that surface heterogeneity destroys the well-structured adsorbent's layering.
Therefore, as it has been previously noted in \cite{goodwin2017mean} the surface heterogeneity makes relatively simple mean-field models more appropriate for the description of the capacities behavior.
}
\amended{
Similar to works \cite{daikhin1996double, daikhin1998nonlinear} we use the perturbation theory to obtain the boundary condition for the Poisson equation \cite{dambrine2016numerical}. But unlike approach \cite{daikhin1996double, daikhin1998nonlinear} we account for the local properties of the random surface to obtain the average ions distributions closing a system of the self-consistent equations.
}

\begin{figure}
    \centering
    \includegraphics[width=0.45\textwidth]{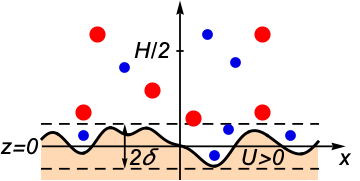}
    \caption{The sketch illustrates the characteristic distribution of the asymmetric ions near rough surface at positive applied potential.}
    \label{fig:sketch}
\end{figure}

In our model, the solid surface profile is the correlated Gauss process $z_s(x)$, where $x$ is the coordinate in the lateral direction, the standard deviation $\delta$ and the correlation length $\lambda$ define the roughness in the normal direction and the characteristic lateral structure, respectively (see Fig.~\ref{fig:sketch}). We put the origin of the normal coordinate $z$ so that the solid profile average equals zero $\overline{z_s(x)}=0$. 
The standard deviation defines the vertical distribution of the solid media and the correlation properties correspond to the decreasing exponential function $\overline{z_s(x)z_s(x+t)}\sim e^{-|t|/\lambda}$. Therefore, the small $\lambda$ induces the palisade of the solid peaks, the large $\lambda$ results in the formation of the sparse structure allowing the fluid molecules to penetrate into perturbed/rough region of the solid medium. Thus, to describe the ions molecules' behavior near rough surfaces, both normal $\delta$ and correlation $\lambda$ parameters are crucial. Such a two-parametric random surface model provides height-profiles mimicking the geometry of the real materials \cite{ciraci2020impact, gadelmawla2002roughness} and has a substantial advantage over deterministic well-structured geometries allowing description randomly distributed surface-heterogeneity (defects and functional groups) \cite{wang2020electrode, evlashin2020role}.

The electrolyte fills the available space that results in the inhomogeneous density distributions of the ions mixture $\rho_i = \rho_i(x,z)$. Accounting for the applied potential $U$ the electrostatic field $\psi  = \psi(x, z)$ inside the pore is defined by the Poisson equation and the boundary conditions
\begin{equation}
\label{eq:Poisson}
\begin{array}{rlrr}
\beta e \Delta \psi &= -4 \pi \lambda_B q & \text{in} & D, \\ 
\psi &= U & \text{at} & \partial D,
\end{array}
\end{equation}
where $\beta = 1 / k_B T$, $k_B$ is Botlzman constant, $T$ is the temperature, $e$ is electron charge, $\Delta = \partial_{xx} + \partial_{zz}$ is the 2D Laplace operator, $q = \sum_i Z_i \rho_i$, $Z_i$ are the valencies, $\lambda_B~=~\beta e^2/(4\pi\epsilon \epsilon_0)$ is the Bjerrum length. The domain $D~=~\{x,z_s(x)<z<H/2\}$ is half-space of the pore of width $H$ above rough surface $\partial D=\{x,z=z_s(x)\}$.
\amended{
We impose the zero field condition at the half width $\psi(H/2)=0$ and, thus, consider sufficiently large pores.
}

Subdividing $D$ into the (bulk) volume $D_v=\{x,\delta<z<H/2\}$ and the (near-)surface $D_s=\{x,-\delta<z<\delta\}$ domains and applying perturbation procedure \cite{dambrine2016numerical} involving matching expansions for the electrostatic potential and charge density with respect to $\varepsilon~=~2\delta / H \ll 1$ in $D_v$ and $D_s$, we show (see \cite{supplementary}, section I) that the average electrostatic field $\overline{\psi} = \overline{\psi}(z)$ above rough surface can be approximated with an error $O(\varepsilon^2)$ by the piece-wise solution
\begin{equation*}
\overline{\psi} = \left\{
\begin{array}{cc}
\overline{\psi}_v, & z \geq \delta \\
\overline{\psi}_s, & z < \delta. 
\end{array}
\right.
\end{equation*}
\amended{
Here the averaging is performed over the realisations of the Gauss random process with exponentially decaying lateral correlation representing the rough surface (please see equations S10 and S11 of \cite{supplementary} for exact definition via one- and two-point distribution functions). 
}
The field in the volume domain $\overline{\psi}_v$ is given by the solution of the following problem
\begin{equation}
\label{eq:internal}
\begin{array}{rlr}
\beta e \partial_{zz} \overline{\psi}_v &= -4 \pi \lambda_B \overline{q}, & z \geq \delta, \\ 
\overline{\psi}_v &= U +\delta \partial_z\overline{\psi}_v, & z = \delta,
\end{array}
\end{equation}
and the field in the surface domain $\overline{\psi}_s$ is given by 
\begin{equation}
\label{eq:external}
\overline{\psi}_s=U+z\partial_z\overline{\psi}_v|_{z=\delta}, z<\delta.
\end{equation}
$\overline{\psi}_s$ is linear in $z$ and naturally equals to applied voltage at the apparent boundary of the pore $z = 0$. The slope of the $\overline{\psi}_s$ dependency on $z$ is defined by the gradient of the electrostatic field $\overline{\psi}_v$ at the boundary $z = \delta$ such that $\overline{\psi}$ is smooth in the entire domain. The problem \eqref{eq:internal} for the average field $\overline{\psi}_v$ is decoupled from the surface domain. Thus, the density distribution in the inner region is only needed to determine the electrostatic field in the entire pore. However, the density in the surface domain crucially influences the charge and capacitance properties. The total charge is defined by the summation of the volume and surface regions $Q=Q_s+Q_v$, where the cumulative contributions are defined from the following integrals $Q_s=\int_{-\delta}^\delta e\overline{q}(z) dz$ and $Q_v=\int_{\delta}^{H/2}\sum e\overline{q}(z) dz$.

Besides the electrostatic forces, the ions interact with the solid boundaries via \amended{Lennard-Jones potential}. We model this interaction as hard wall repulsion and consider the ionic liquid as hard spheres mixture \cite{hartel2017structure}. 
In the electrodes with flat surfaces, the minimal distance between ions and solid boundaries is defined by hard spheres radii $d_i/2$. This behavior contrasts with the molecules distributions near rough surfaces that allow ions to penetrate into the rough region of solid medium. Therefore, the function of the ions' spatial distribution starts from some points $z_i<d_i/2$.  This starting points $z_i$ may be negative for extremely rough surfaces and tends to hard walls value $d_i/2$ as the roughness becomes insignificant. Considering the rough surface as the Gauss correlated process the contact conditions can be calculated as functions of the relative roughness parameters $z_i(\delta/d_i,\lambda/d_i)$ \cite{supplementary}. The details of the dependence of ions penetration into solid media on the surface roughness and the diameter of molecules can be found in \cite{supplementary}, section II. In comparison with the flat surface, the roughness results in the region which is filled by both ions and solid molecules, 
but at the same time, the vertical distribution of the solid media decreases the space permitted for the ions. It can be accounted by excluding the ratio of solid media at each level $z$ from the whole covered surface, then the permitted surface area as a function of $z$ has the following form:
\begin{equation}
\label{eq:rough_surface}
S(z)=S_0 s(z)=S_0\left(1-\frac{1}{2}\erfc \frac{z}{\sqrt{2}\delta}\right)
\end{equation}
where $S_0$ is the area of the surface projection on the lateral plane. Expression \eqref{eq:rough_surface} depends on the standard deviation only and defines the vertical impact of the roughness (see derivation in \cite{supplementary}, section II). 

To describe the electrolyte near rough surfaces we take into account the average properties of the random rough surface: the averaged electrostatic field; the modified configuration volume; the hard shere interaction with the rough surface. 
\amended{
We account for these effects in terms of the Helmholtz free energy potential $F$ defined in the volume free from solid media $\int S(z) dz$
\begin{equation}
    \label{eq:Helmholtz_free_energy_1}
    F[\{\overline{\rho}_i\}]=\int S(z) dz \sum_i\left[U^\text{ext}_i(z)+\overline{\psi}(z) Z_i\right]\overline{\rho}_i(z)+F^{HS},
\end{equation}
}
$U^\text{ext}_i(z)$ is the hard boundary potential,
\amended{
and $F^{HS}$ is the contribution from hard spheres interaction. 
The equilibrium condition defines the chemical potentials $\mu_i=\frac{1}{S(z)}\frac{\delta F[\{\overline{\rho}_i\}]}{\delta \overline{\rho}_i}$, which are constant across the volume.
}
As one can see from the detailed calculations in \cite{supplementary}, section III the density distributions have the following form: 
\begin{equation}
    \label{eq:densities_1}
    \overline{\rho}_i(z)=s(z)\theta(z-z_i)\rho^0_i\frac{e^{-Z_i\beta e \overline{\psi}}}{1-\sum_i\gamma_i+\sum_i\gamma_i e^{-Z_i\beta e \overline{\psi}}}
\end{equation}
where $\rho^0_i$ is the bulk density describing the component far enough from the surface, $\gamma_i=v_i\rho_i^0$ are the model parameters showing the packing density of the fluids. It is worth noting that in the case of $z_i=0$, $\rho^0_i=\rho^0$ and $v_i=v$ the result \eqref{eq:densities_1} agrees with work \cite{kornyshev2007double} for $\gamma=\sum_i v_i\rho^0_i$.


\amended{In contrast with the perturbation theory for Poisson-Boltzmann equation \cite{daikhin1996double, daikhin1998nonlinear} where the calculation of the electrostatic field corrections up to the second order in roughness has been required to determine the first nonvanishing correction to capacity, we keep only first order terms while deriving the system \eqref{eq:internal}, \eqref{eq:external} (see \cite{supplementary}, section I). Notable result of our calculations, illustrated below, is that  in the case of non-equal minimal distances between the ions' centers and solid surface $z_i$ the first order theory is sufficient to reproduce non-trivial contribution of roughness to differential capacity.}




We consider the binary mixture of the hard sphere molecules with the opposite charges $Z_1=-Z_2=1$ and non-equal diameters $d_1\neq d_2$. In the absence of applied voltage, the mixture is electrically neutral and composition bulk densities are equal $\rho^0=\rho^0_i$.
The following dimensionless variables are introduced $U^*=e U/k_B T$, $z^*=z/d_m$, $H^*=H/d_m$, $Q^*=Q /\rho^0 d_m$, $\lambda_B^*=\lambda_B d_m^2 \rho^0$, where $d_m = \min(d_1, d_2)$ is the molecular diameter of the smallest component.

\begin{figure*}
    \centering
    \includegraphics[width= \textwidth]{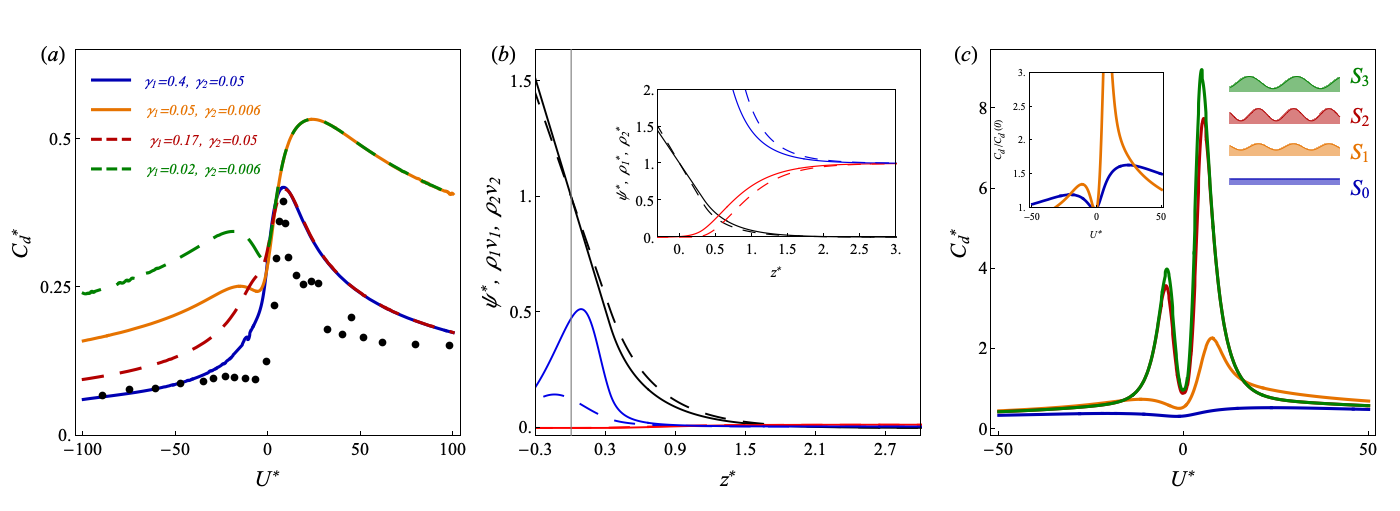}
    \caption{
    \label{fig:dif_cap_flat}(a): Flat electrode DC of ions mixtures with $d_1=2d_2$ (solid lines) and $d_1=1.5 d_2$ (dashed lines). MD simulations results \cite{fedorov2008ionic} also shown for reference (disks).
    \label{fig:potential_rough}(b): The relative electrostatic potential $\psi^*/U^*$ (black), cations $\rho_1 v_1$ (red) and anions $\rho_2 v_2$ (blue) packing density distributions at two applied potentials $U^*=4$ (solid lines) and $U^*=8$ (dashed lines). Molecular diameters ratio $d_1=4/3 d_2$, $\gamma_1=0.014$, $\gamma_2=0.006$, $\lambda_B^*=0.25$. Surface parameters ($\delta^*=0.33$, $\lambda^*=1.66$, $z_1^*=-0.17$, $z_2^*=-0.31$) correspond to the surface S$_3$ shown in Fig.~\ref{fig:DC_rough}(c). Inset shows the detailed distribution inside inner region using normalized density $\rho_i^*=\rho_i/\rho_0$.
    \label{fig:DC_rough}(c): The DC for the binary mixture with $d_1=4/3 d_2$  with $\gamma_1=0.014$, $\gamma_2=0.006$, $\lambda_B^*=0.25$ near the various surface geometries: from flat S$_0$ to significantly rough S$_3$ (the surface parameters can be found in Supporting Information, section 2). Inset shows the larger scale of DC for flat S$_0$ and slightly rough S$_1$ surfaces.
    }
    \label{fig:fig2abc}
\end{figure*}

First, we investigate the case of a flat pore wall surface $\delta = 0$. The minimal distances between the flat surface and the center of ion are equal to ions' radii $z^0_i=d_i/2$. 
Fig.~\ref{fig:dif_cap_flat}(a) 
shows the dimensionless DC $C_d=\partial Q^*/\partial U^*$ as a function of the potential $U^*$ for the case of cations larger than anions. Similarly to the model \cite{kornyshev2007double}, the high potential limit of the DC is $C_d\sim 1/\sqrt{\gamma_2 |U|}$ and $C_d\sim 1/\sqrt{\gamma_1 |U|}$ for positive and negative $U$, respectively. Since composition bulk densities $\rho^0=\rho^0_i$ are equal for $Z_1=-Z_2=1$, the ratio of right and left wings of $C_d$ shown in Fig.~\ref{fig:dif_cap_flat}(a) is defined by the ions sizes as $(\gamma_1/\gamma_2)^{1/2}=(d_1/d_2)^{3/2}$. As one can see from Fig.~\ref{fig:dif_cap_flat}(a) the DC at negative potential, where a contribution of the larger cations to charge dominates, is lower than at a positive one. 
Such asymmetric behaviour agrees with published data of MD simulations \cite{fedorov2008ionic} shown in Fig.~\ref{fig:dif_cap_flat}(a).
Also,  Fig.~\ref{fig:dif_cap_flat}(a) demonstrates that the number of capacitance maxima depends on the bulk density $\rho^0$. The curves corresponding to sufficiently low $\gamma_i$ exhibit two maximum points — the sharp and diffuse peaks at regions of small and large ions prevailing contribution to the total charge, respectively. 



The roughness induces more striking changes in the capacitance properties. To isolate the impact from the surface geometry, we considered slightly asymmetrical electrolyte with the following molecular diameters ratio $d_2=4/3 d_1$. 
\amended{
The characteristic examples of the calculated electrostatic fields and ions distributions as functions of the coordinate $z$ are shown in Fig.~\ref{fig:potential_rough}(b).
As one can see from Fig.~\ref{fig:potential_rough}(b) the considered parameters allow us to apply our approach for pores larger than six molecular diameters, that for ionic liquids corresponds to mesopores $H>\SI{2}{nm}$.  
}
The rough surface allows molecules to reach the surface region ($z < \delta$), where the electrostatic field is defined by \eqref{eq:external}. 
Inside the surface region, the absolute value of electrostatic field $|\psi^*(z)|$ locally increases (see Fig.~\ref{fig:potential_rough}(b)) due to the sharp decrease of the permitted surface $S(z)$. Such potential behavior crucially influences the ions distributions improving the spatial separations of the co-/counter-ions. Fig.~\ref{fig:potential_rough}(b) shows that the counter-ions cumulative effect from the surface region becomes overwhelming as the applied potential increases, while the electrolyte behavior in the volume domain ($z>\delta$) remains almost unperturbed. The contribution of $Q_s$ to the total charge is notable and significantly improves the capacity properties.
\amended{
Therefore, the surface roughness induces the enhancement of the integral capacity that is in agreement with published simulations, for example, \cite{xing2012nanopatterning, vatamanu2015non} and recent experiment \cite{wei2020surface}.
}
Moreover, in work \cite{xing2012nanopatterning} the capacitance enhancement has been also attributed to the local increase of the electrostatic field near rough surface and its influence co-/counter-ions distribution inside electrode-electrolyte interface (cumulative density). 
  
To describe the total charge dependency on the applied potential, we calculated the DC $C_d$ for the rough electrodes. Fig.~\ref{fig:DC_rough}(c) shows our results for various surface morphology varying from flat to significantly rough. As one can see from Fig.~\ref{fig:DC_rough}(c) the roughness notably increases the capacitance and, in particular, the values of $C_d(U)$ maxima. Indeed, the quantitative comparison of the results from Fig.~\ref{fig:DC_rough}(c) shows that the flat DC is almost constant while the surfaces with the higher roughness exhibit larger $C_d$ values and extremely sharp peaks. Such DC behaviour agrees with the observations from MD simulations
\cite{vatamanu2011influence, xing2012nanopatterning} for ionic liquids inside rough electrodes. 
The calculated DCs curves from Fig.~\ref{fig:DC_rough}(c) conserve the number of peaks, while the MD simulations \cite{vatamanu2011influence,xing2012nanopatterning} predict the roughness induced appearance/disappearance of the DCs curves maxima. The results of work \cite{xing2012nanopatterning} allow us to explain this discrepancy. The authors of \cite{xing2012nanopatterning} demonstrated that the cumulative center-of-mass ion densities describe DCs curve at a finite range of the potential containing only two peaks and the formation of the additional peaks is related to the steric effect of the ions spatial orientations. Therefore, our model's application is limited by the influences of the electrostatic potential and ion's distributions of center of mass near rough electrodes. 
However, the roughness induced transition from two-peaks to one-peak DC curve simulated in work \cite{vatamanu2011influence} can be qualitatively described in terms of our model considering the suppressed maximum instead of full extinction. 
Let us compare the ideal flat geometry and the surface with the lowest roughness, which are noted as S$_0$ and S$_1$ in Fig.~\ref{fig:DC_rough}(c). The surface region of S$_1$ is filled by the smallest ions mainly, leaving the other component at the volume region. Such asymmetry promotes only one enhanced peak corresponding to the situation when the smallest-size component is counter-ions. Thus, as one can see from inset in Fig.~\ref{fig:DC_rough}(c) the increase of the surface roughness transforms two comparable peaks S$_0$ of DC to one dominating peak S$_1$.

\amended{
The quantitative predictions for the concentrated ionic liquids presented above are limited by the simple mean-field approach used in the chemical potentials calculations. This model omits the effects from both the electrostatic correlations \cite{hartel2017structure} and accurate hard-sphere equation of state depending on the weighted densities \cite{roth2010fundamental}. However, it is possible to address these shortcomings extending the proposed approach for the more sophisticated models beyond the mean-field ones, for example, the Bazant-Storey-Kornyshev (BSK) theory \cite{bazant2011double} accounting for the electrostatic correlations and very recent work \cite{de2020interfacial} showing the spatial oscillations of the ions density.
}
\amended{
Since in the BSK theory \cite{bazant2011double} the calculation of the ions chemical potentials is very similar to the model considered here \cite{kornyshev2007double}, the rough surface results from the current work are applicable to BSK theory as well. More recent model \cite{de2020interfacial} is formulated in terms of the weighted charge densities, that results in the Helmholtz free energy functional similar to FMT \cite{roth2010fundamental}. The random surface extension of the FMT based functional was already investigated in the problem of uncharged molecules adsorptions \cite{aslyamov2017density, aslyamov2019random}. Therefore, our approach applied to both models \cite{bazant2011double, de2020interfacial} can provide equations for the ions density distributions near rough surfaces, which will depend on the averaged electrostatic fields. The electrostatic properties can be defined separately, considering the corresponding differential equations with the boundary conditions on the rough (random) surfaces. In model \cite{de2020interfacial} the electrostatic field is defined by the Poisson equation with the weighted charged density in the right hand side. Therefore, the application of the perturbation theory \cite{dambrine2016numerical} will result in the expressions similar to the current work. Also, the perturbation technique \cite{dambrine2016numerical} could be used for the modified Poisson equation from \cite{bazant2011double}, that potentially will demand a higher order expansion. 
Thus, similarly to current work, separated rough surface modifications for the ions distribution densities and electrostatic fields will provide the self-consistent equations for models \cite{bazant2011double, de2020interfacial}.
}

In future works, besides models \cite{bazant2011double, de2020interfacial} we will mplement the rough surface approach into classical Density Functional Theory (c-DFT) which successfully describes static \cite{hartel2017structure} and dynamic \cite{aslyamov2020relation} properties of the supercapacitors. Despite several versions of c-DFT describes the adsorption of the neutral molecules on the rough uncharged surfaces \cite{neimark2009quenched, aslyamov2017density, aslyamov2019random} the electrostatic c-DFT has been previously applied to flat electrodes only. Detailed theoretical investigation of the electrolyte behaviour near a rough surface could explain the difference of the roughness impact on the ionic liquids and the solutions described in \cite{vatamanu2017charge} accounting for the the solvent molecules compatible adsorption which effectively decreases the surface roughness.

\amended{
One of the interesting features of our theory is the relative character of the roughness influence. Indeed, as one can see from the model description shown here and in works \cite{khlyupin2017random, aslyamov2017density}, the averaged properties are defined by the dimensionless parameters $(\delta/d_i, \lambda/d_i)$. Therefore, it is possible to investigate the roughness-induced effects considering only one surface sample and a set of cations/anions with various diameters. This scheme is similar with Parsons-Zobel plot \cite{parsons1965interphase} showing the dependence of the inverse value of the experimentally measured capacity $1/C_\text{exp}$ on the theoretically predicted one $1/C_\text{th}$ for the different electrolytes. 
Regarding the process of the experimental measurements, the impedance data of non-ideal capacitors is often interpreted in terms of a constant phase element \cite{lockett2008differential}. This approach provides the power-law dependence of the capacitance on the frequency $\omega$ in the form $C\sim (\omega i)^{\alpha-1}$, where $\alpha$ is the system parameter $0<\alpha\leq 1$. The experiments show that $\alpha\to 1$ as more smoother and cleaner electrodes are considered \cite{lockett2008differential}. The connection between the frequency dependence of the capacity and roughness has been identified a long time ago in work \cite{borisova1950determination}. Since the porous materials roughness is the multi-scale characteristic, it is a complicated problem to identify the explicit origin of the observed frequency-dependence.  
Studies \cite{kerner1998impedance, kerner2000origin} revealed that it is the atomic scales surface heterogeneities, which induce the dispersion behavior.
Our model accounts for such scale of the heterogeneity, that can be used to develop the rough surface dynamics model describing the ions adsorption at time-dependent potentials $U=U_0 \cos \omega t$. For example, the impedance for flat electrodes can be calculated using the dynamic density functional theory \cite{babel2018impedance}. Thus, random surface extension of the electrolyte c-DFT approach could be useful for the investigation of the capacitance dispersion.   
}

In conclusion, we developed a theory describing the ions distribution structure and accounting for the realistic roughness of the porous electrodes. 
Our model predicts the significant capacitance increase induced by ions-scale roughness. 
Moreover, we observed that the shape of the DC dependency on applied potential changes notably with a variation of roughness and becomes sharper as the roughness increases. 

\begin{acknowledgments}
T.A. acknowledges the financial support from the Russian Science Foundation (project number: 20-72-00183). K.S. is grateful to Schlumberger management for the permission to publish this work. The authors are grateful to Mikhail Stukan for useful comments. 
T.A. and K.S. contributed equally to this work.
\end{acknowledgments}
\newpage
\appendix
\onecolumngrid
\renewcommand\thefigure{S\arabic{figure}}
\renewcommand\theequation{S\arabic{equation}}

\section{Electrostatic field near rough surface}
Derivation of the system (2), 
(3) 
relating the ensemble-averaged electrostatic field and charge density distributions near rough surface from the system (1) 
formulated in terms of a single realization of the surface can be divided in two principle steps. First, one can derive a series of boundary value problems for the deterministic but rough surface using asymptotic expansions with respect to the small parameter characterizing the roughness. At this step we follow the approach presented in \cite{dambrine2016numerical} and outline derivation below. Second, the boundary value problems can be formally averaged over ensemble and used to obtain the problem for the approximation of ensemble-averaged field.

In order to employ asymptotic expansion with respect to the surface roughness we recast equations (1) 
to dimensionless form using the pore width $H$ and the applied voltage $U$ as a length and potential scales respectively. The charge density scale is then $\beta e U / 4 \pi \lambda_B H^2$. The dimensionless Poisson equation is given by 
\begin{equation}
\label{eq:Poisson_dimensionless}
\begin{array}{rlrr}
\Delta \psi &= -q & \text{in} & D, \\ 
\psi &= 1 & \text{at} & \partial D,
\end{array}
\end{equation}
the domain $D = \{x, z_s(x) < z < 1/2\}$ and its boundary $\partial D = \{x, z = z_s(x)\}$.

Next, we decompose the domain $D$ with deterministic but corrugated boundary $\partial D$ into two subdomains $D~=~D_v~\cup~D_s$ divided by a smooth artificial boundary. Taking into account that realizations of Gaussian random process used here to model the pore surface mostly lie in the interval $-\delta < z < \delta$, we specify the artificial boundary $\partial D_v = \{x, z = \delta\}$. Thus, the (bulk) volume (referred to as internal in terminology of  \cite{dambrine2016numerical}) domain $D_v = \{x, \delta < z < 1 / 2\}$, the complementing (near-)surface (referred to as external in terminology of \cite{dambrine2016numerical}) domain $D_s = \{x, z_s (x) < z < \delta\}$ and its outer boundary $\partial D_s = \{x, z = z_s(x)\}$.

Using the rough surface shape $z = z_s(x)$ one can introduce the characteristic size of the surface domain $\varepsilon = 2 \delta$, its scaled width $h(x) = (\delta - z_s(x)) / \varepsilon$ and define the coordinate $\zeta = (\delta - z) / \varepsilon h$ in $D_s$ such that $D_s~=~\{x, 0 < \zeta < 1\}$ and the boundaries $\partial D_v = \{x, \zeta = 0\}$, $\partial D_s = \{x, \zeta = 1\}$. Following \cite{caloz2006asymptotic, dambrine2016numerical} we adopt double expansions for the electrostatic potential and charge density with respect to $\varepsilon$ using original coordinates $x, z$ in $D_v$ and scaled coordinates $x, \zeta$ in $D_s$
\begin{equation}
\label{eq:expansion}
u = 
\left\{
\begin{array}{ccc}
u_v \overset{\text{def}}{=} \sum\limits_{k = 0}^{\infty} \varepsilon^k u_{v, k} (x, z) & \text{in} & D_v, \\
u_s \overset{\text{def}}{=} \sum\limits_{k = 0}^{\infty} \varepsilon^k u_{s, k} (x, \zeta) & \text{in} & D_s,
\end{array}
\right.
\end{equation}
where $u = \{\psi, q\}$. The problem \eqref{eq:Poisson_dimensionless} is reformulated as a transmission problem coupling both values $\psi$ and normal gradients $\partial_{\mathbf{n}} \psi$ of the volume and surface solutions for electrostatic field at the boundary $\partial D_v$
\begin{equation}
\label{eq:transmission_problem}
\begin{array}{rlrr}
\Delta \psi_v &= -q_v & \text{in} & D_v, \\ 
\psi_v &= \psi_s & \text{at} & \partial D_v, \\
\partial_{\mathbf{n}} \psi_v &= \partial_{\mathbf{n}} \psi_s & \text{at} & \partial D_v, \\
\Delta \psi_s &= -q_s & \text{in} & D_s, \\
\psi_s &= 1 & \text{at} & \partial D_s.
\end{array}
\end{equation}

Rewriting the equations \eqref{eq:transmission_problem} for the surface domain problem with respect to $x, \zeta$, substituting expansions \eqref{eq:expansion} and matching the coefficients of the terms with the same power of $\varepsilon$ one can obtain

$O\left(1\right)$ {\it problem}:
\begin{equation*}
\label{eq:transmission_problem_0}
\begin{array}{rlrr}
\Delta \psi_{v, 0} &= -q_{v, 0} & \text{in} & D_v, \\ 
\psi_{v, 0} &= \psi_{s, 0} & \text{at} & \partial D_v, \\
0 &= \partial_{\zeta} \psi_{s, 0} & \text{at} & \partial D_v, \\
\partial_{\zeta\zeta} \psi_{s, 0} &= 0 & \text{in} & D_s, \\
\psi_{s, 0} &= 1 & \text{at} & \partial D_s.
\end{array}
\end{equation*}

Explicit solution for the surface domain is given by 
\begin{equation}
\label{eq:transmission_problem_0_ext}
\psi_{s, 0} = 1
\end{equation}
and thus, due to the coupling conditions at $\partial D_v$, $\psi_{v, 0}$ solves the boundary value problem with the potential prescribed at the shifted boundary $z = \delta$
\begin{equation}
\label{eq:transmission_problem_0_int}
\begin{array}{rlrr}
\Delta \psi_{v, 0} &= -q_{v, 0} & \text{in} & D_v, \\ 
\psi_{v, 0} &= 1 & \text{at} & \partial D_v.
\end{array}
\end{equation}

$O\left(\varepsilon\right)$ {\it problem}:
\begin{equation*}
\label{eq:transmission_problem_1}
\begin{array}{rlrr}
\Delta \psi_{v, 1} &= -q_{v, 1} & \text{in} & D_v, \\ 
\psi_{v, 1} &= \psi_{s, 1} & \text{at} & \partial D_v, \\
-h \partial_z \psi_{v, 0} &= \partial_{\zeta} \psi_{s, 1} & \text{at} & \partial D_v, \\
\partial_{\zeta\zeta} \psi_{s, 1} &= 0 & \text{in} & D_s, \\
\psi_{s, 1} &= 0 & \text{at} & \partial D_s.
\end{array}
\end{equation*}

Solution for the surface domain in this case is
\begin{equation}
\label{eq:transmission_problem_1_ext}
\psi_{s, 1} = -(\zeta - 1) h \partial_z \psi_{v, 0}|_{z = \delta}
\end{equation}
and the coupling conditions at $\partial D_v$ imply that $\psi_{v, 1}$ solves the boundary value problem
\begin{equation}
\label{eq:transmission_problem_1_int}
\begin{array}{rlrr}
\Delta \psi_{v, 1} &= -q_{v, 1} & \text{in} & D_v, \\ 
\psi_{v, 1} &= h \partial_z \psi_{v, 0} & \text{at} & \partial D_v.
\end{array}
\end{equation}

Although the procedure can be continued for higher orders of $\varepsilon$ resulting in more accurate approximation for $\psi$ \cite{dambrine2016numerical}, we restrict consideration to $O(\varepsilon)$ problem and proceed with averaging over ensemble of random surfaces. To perform averaging we note that the equations \eqref{eq:transmission_problem_0_int} don't involve the scaled surface domain width $h$. Accordingly, the ensemble averages $\overline{\psi}_{v,0}~=~\psi_{v,0}$ and $\overline{h \partial_z \psi_{v, 0}}~=~\overline{h} \partial_z \overline{\psi}_{v, 0}$. Using the latter equality one can apply ensemble averaging to both sides of the equations \eqref{eq:transmission_problem_0_int}, \eqref{eq:transmission_problem_1_int} and combine the resulting equations 
to get
\begin{equation}
\label{eq:internal_dimensionless}
\begin{array}{rlrr}
\Delta \overline{\psi}^{[1]}_v &= -\overline{q}^{[1]}_v & \text{in} & D_v, \\ 
\overline{\psi}^{[1]}_v - \delta \partial_z \overline{\psi}^{[1]}_v&= 1 & \text{at} & \partial D_v
\end{array}
\end{equation}
for the partial sums $\overline{\psi}^{[1]}_v = \overline{\psi}_{v,0} + \varepsilon \overline{\psi}_{v,1}$ and $\overline{q}^{[1]}_v = \overline{q}_{v,0} + \varepsilon \overline{q}_{v,1}$. Here we also used that $\overline{z_s}  = 0$ and accordingly $\overline{h} = 1/2$.

Similarly, using the equations \eqref{eq:transmission_problem_0_ext}, \eqref{eq:transmission_problem_1_ext} and neglecting higher order terms we find that the $\overline{\psi}^{[1]}_s = \overline{\psi}_{s,0} + \varepsilon \overline{\psi}_{s,1}$ satisfies
\begin{equation}
\label{eq:external_dimensionless}
\overline{\psi}^{[1]}_s = 1 - (\zeta - 1) \varepsilon \overline{h} \partial_z \overline{\psi}^{[1]}_v|_{z = \delta} = 1 + z \partial_z \overline{\psi}^{[1]}_v|_{z = \delta}.
\end{equation}

The solution of \eqref{eq:internal_dimensionless}, \eqref{eq:external_dimensionless} approximate the mean electrostatic potential in volume and surface domains with the first order accuracy $\overline{\psi} = \overline{\psi}^{[1]} + O(\varepsilon^2)$. Dropping superscripts from \eqref{eq:internal_dimensionless}, \eqref{eq:external_dimensionless}, restoring the dimensional variables and noting that for homogeneously rough surfaces $\overline{\psi} = \overline{\psi}(z)$, we readily arrive at the problem (2), (3).
\section{Random surface model and hard sphere contact condition}
Here we briefly discuss the local geometrical properties of the correlated random process which models the realistic surface roughness. The detailed derivation can be found in work \cite{khlyupin2017random} dedicated to the effective molecular potential between a fluid molecule and the solid media with the rough surface. In this approach the properties of the rough surface geometry corresponds to the Gauss correlated random process defined by the following one- and two-points distribution functions:
\begin{equation}
\label{eq:one-point}
w_1(z)=\frac{1}{\sqrt{2\pi\delta^2}}\exp\left(-\frac{z^2}{2\delta^2}\right),
\end{equation}
\begin{equation}
\label{eq:two-point}
w_2(z_1,x;z_2,x+t)=\frac{1}{2\pi\sigma^2\sqrt{1-K(t)^2}}\exp\left(-\frac{z_1^2+z_2^2-2K(t)z_1z_2}{2\sigma^2(1-K(t)^2)}\right),
\end{equation}
where $K(s)$ is the lateral correlation function, $\delta$ is the standard deviation. Therefore, the rough solid heigth profile $z_s$ can be considered as the realisations of such random process. More precisely we consider the random profiles $z_s$ with the zero average and exponentially decreasing lateral correlations: 
\begin{equation}
\label{eq:appendix_average_correlation}
\overline{z_s(x)}=0, \quad \quad \overline{z_s(x)z_s(x+t)}=K(s)= e^{-|t|/\lambda},
\end{equation}
where over-line symbol is the average over the random process, $\lambda$ is the correlation length. Thus, this approach allows us to consider two-dimensional rough surface in terms of two parameters $\delta$ and $\lambda$ describing the roughness in the normal direction and the lateral geometry, respectively. 

\begin{figure}[b]
    \centering
    \includegraphics[width=\textwidth]{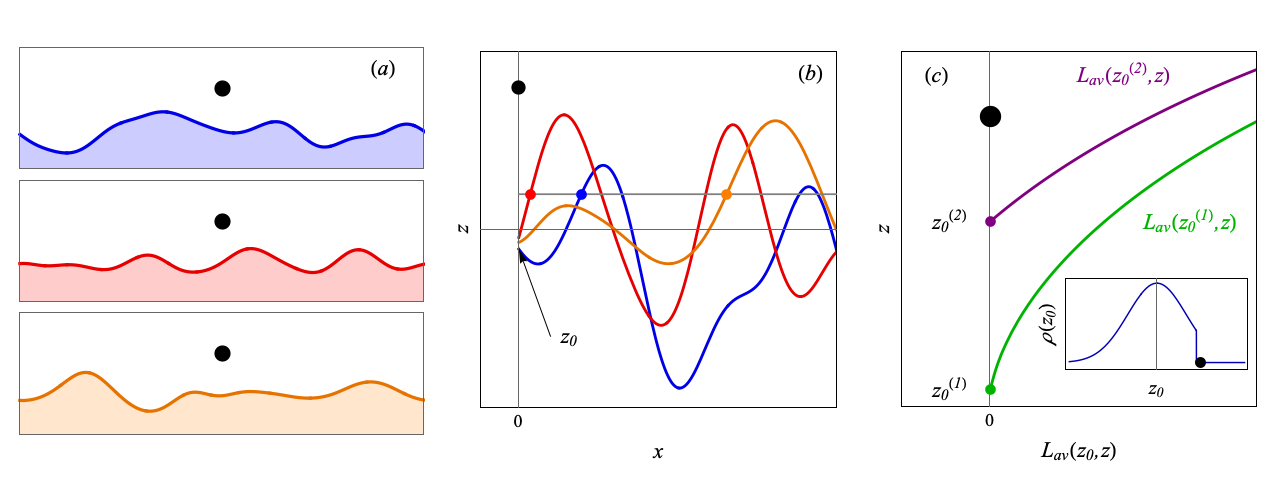}
    \caption{The first step of the random surface averaging for the fluid molecule (black dot) at point $(0,z_f)$:
    (a) the characteristic realisations of the random process with fixed parameters $\delta$ and $\lambda$, (b) the colored curves correspond to the surface profiles $z_s(x)$ which pass through the point (0, $z_0$), the colored dots define the moments when the profiles exhibit the first crossing of the certain level $z$. (c) The colored curves (equation X) are the averaged lengths corresponding to "the first passage time" for the random profiles starting from the points $z_0^{(1)}$ and $z_0^{(2)}$.}
    \label{fig:RS_fig_1}
\end{figure}

We consider spherical fluid molecule with the diameter $d$ located near rough surfaces at the point $(0, z_f)$ (the $x$-axes origin is the location of the fluid molecule). In this case we take into account the process realisation passing below the fluid location that approximately equivalent the condition $z_s(0)\leq z_f-d/2$. One can account for the proper realisations considering the profiles which pass through the certain starting point $(0, z_0)$ for all $z_0\leq z_f-d/2$.  Fig.~\ref{fig:RS_fig_1}(b) illustrates such profiles which can be described by "the first passage time (FPT) model" considering the $x$-coordinate instead of time. 
Indeed the random process starts at moment $x=0$ and the crosses some level $z$ as the value of $x$-coordinate increases Fig.~\ref{fig:RS_fig_1}(b). Therefore, the averaging of $z$-level first hitting defines the average length along the $x$-coordinate $L_\text{FPT}$.
In work \cite{khlyupin2017random} the authors calculated the analytical expression for $L_\text{FPT}(z_0,z)$ for the Markov random correlated process:
\begin{equation}
    \label{eq:appendix_L_aver_step_1}
    L_\text{FPT}(z_0,z)=\frac{\lambda}{\delta^2}\int_{z_0}^z e^{\frac{\xi^2}{2\delta^2}}d\xi\int_{-\infty}^\xi e^{-\frac{\eta^2}{2\delta^2}}d\eta=\lambda\int_0^1d\tau\frac{e^{\frac{z^2}{2\delta^2}(1-\tau^2)}-e^{\frac{z_0^2}{2\delta^2}(1-\tau^2)}}{1-\tau^2}+\frac{\pi}{2}\left(\erfi \frac{z}{\sqrt{2}\delta}-\erfi \frac{z_0}{\sqrt{2}\delta}\right).
\end{equation}
As one can see from expression \eqref{eq:appendix_L_aver_step_1} and Fig.~\ref{fig:RS_fig_1}(c) the length $L_\text{FPT}(z_0, z)$ crucially depends on the starting point $z_0$. The inset in Fig.~\ref{fig:RS_fig_1}(c) shows the probability distribution for the profiles at point $x=0$ accounting for the fluid molecules location $z_f$:
\begin{equation}
    \label{eq:appendix_probability_distribution}
    \rho(z_0)=\frac{w_1(z_0)\theta(z_0-z_f+d/2)}{\int_{-\infty}^{\infty}w_1(z_0)\theta(z_0-z_f+d/2)dz_0}=\frac{w_1(z_0)\theta(z_0-z_f+d/2)}{\int_{-\infty}^{z_f-d/2}w_1(z_0)dz_0}.
\end{equation}
To account for all random process realisations we calculate the average for the length \eqref{eq:appendix_L_aver_step_1} over the starting point $z_0$ distribution \eqref{eq:appendix_probability_distribution}. The result of the averaging at point $z$ depends on the relative position in respect to the fluid molecule's boundary $z_f-d/2$ and can be written as:

\begin{equation}
    \label{eq:appendix_L_1}
    L_{av}(z_f,z)=
    \left\{
    \begin{array}{cc}
         & L_{av}^u(z_f,z), \quad z>z_f-d/2 \\
         & L_{av}^b(z_f,z), \quad z\le z_f-d/2
    \end{array}
    \right.
\end{equation}
In the upper region ($z>z_f-d/2$) the averaging accounts for all starting points in the accordance with the probability distribution \eqref{eq:appendix_probability_distribution}:
\begin{equation}
    \label{eq:appendix_L_up}
    L^u_{av}(z_f,z)=\frac{\int_{-\infty}^{z_f-d/2} L_\text{FPT}(z_0,z) w_1(z_0)dz_0}{\int_{-\infty}^{z_f-d/2}w_1(z_0)dz_0}.
\end{equation}
The bottom region induces an additional constraint $z_0<z<z_f-d/2$:
\begin{equation}
    \label{eq:appendix_L_bot}
    L^b_{av}(z_f,z)=\frac{\int_{-\infty}^{z} L_\text{FPT}(z_0,z) w_1(z_0)dz_0}{\int_{-\infty}^{z_f-d/2}w_1(z_0)dz_0}.
\end{equation}

\begin{figure}[b]
    \centering
    \includegraphics[width=\textwidth]{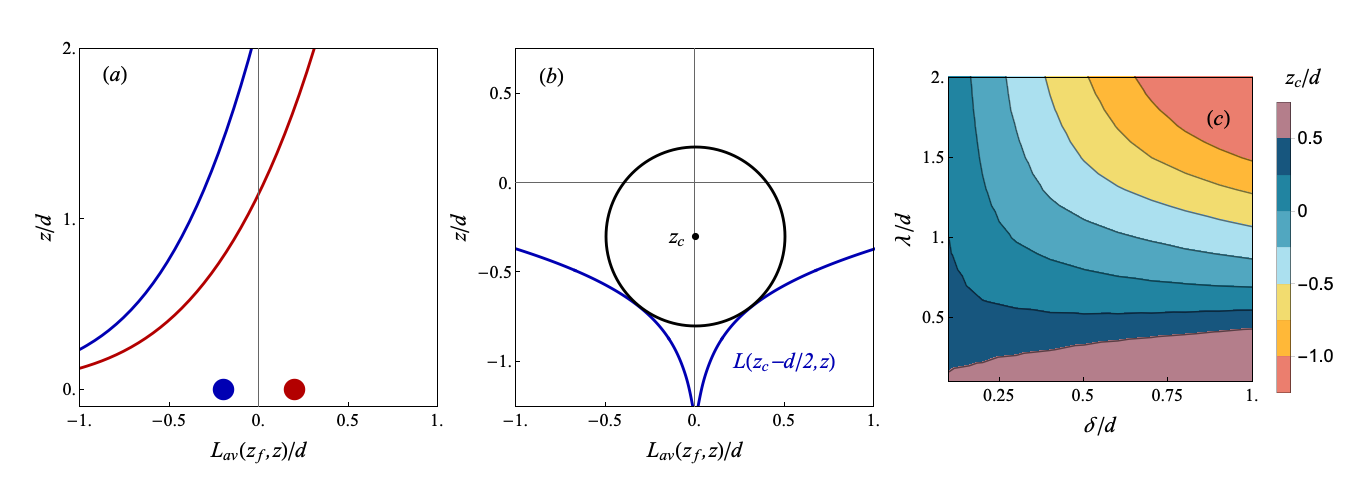}
    \caption{(a) The average length $L_{av}$ (red and blue curves) calculated for various locations of the fluid molecules (red and blue dots). (b) Hard sphere contact condition defining the minimal distance between the fluid and solid media $z_c$.}
    \label{fig:RS_fig_2}
\end{figure}

The average length $L_{av}$ reflects the local properties of the solid surface geometry and depends on the fluid molecules location Fig.~\ref{fig:RS_fig_2}(a). Therefore, the function $L_{av}(z_f, z)$ defines the averaged boundary of the solid media in the vicinity of the fluid molecule at point $z_f$. In work \cite{khlyupin2017random} the authors used this geometrical construction for the integration of the fluid-solid molecular potential. Here, since we consider hard-sphere solid-fluid potential the average interaction corresponds to the contact condition shown in Fig.~\ref{fig:RS_fig_2}(b). As one can see the from Fig.~\ref{fig:RS_fig_2}(b) the molecule with diameter $d$ touches the upper average length (solid boundary) in some contact point. This contact condition defines the starting point $z_c$ of the fluid density distribution near the rough surface with known parameters $\delta$ and $\lambda$. 

We calculate numerically the contact points $z_c(\delta/d,\lambda/d)$ as a function of the dimensionless parameters  $\delta/d$ and $\lambda/d$. The contour plot of this calculations is shown in Fig.~\ref{fig:RS_fig_2}(c).
The nonlinear behavior of  $ z_c(\delta/d,\lambda/d)$ from Fig.~\ref{fig:RS_fig_2}(c) demonstrates that the penetration inside solid crucially depends on the molecular size $d$. We use this rough surfaces property to calculate the starting points $z_i$ for non-symmetric electrolyte (the hard spheres with the diameters $d_i$). For example, the parameters $ z_i(\delta/d_i,\lambda/d_i)$ used in calculations of differential capacity shown in Fig. 4 can be found in Table~\ref{tab:table_1}.
\begin{table}
\centering
\begin{tabular}{|c|c|c|c|c|c|c|} 
 \hline
\# & $d_1^*$ & $d_2^*$ & $\delta^*$ & $\lambda^*$ & $z_1^*$ & $z_2^*$ \\ 
 \hline
S$_1$ &1.33 & 1.00 & 0.17 & 1.23 & 0.28 & 0.10 \\ 
S$_2$ &1.33 & 1.00 & 0.33 & 1.23 & -0.09 & -0.16  \\ 
S$_3$ &1.33 & 1.00 & 0.33 & 1.66 & -0.17 & -0.31  \\ 
 \hline
\end{tabular}
\caption{The geometrical parameters of the considered fluids and surfaces.}
 \label{tab:table_1}
\end{table}

To calculate the permitted surface area $S(z)$ as a function of $z$ we define the ratio of the solid media above level $z$ using the properties of the Gauss random process \eqref{eq:one-point}. Then the surface area permitted for the fluid molecules is defined as the residue after the excluding the ratio of the area occupied by the solids media:
One can obtain this value from the following expression:
\begin{equation}
    \label{eq:appendix_S_1}
    S(z)=S_0\left(1-\int_z^{\infty}w_1(\eta)d\eta\right)=S_0\left(1-\frac{1}{2}\erfc \frac{z}{\sqrt{2}\delta}\right).
\end{equation}
\section{The average distributions of ions near rough surface}
Here we consider the ions near rough surface accounting for the average electrostatic potential $\overline{\psi}(z)$ and averaged properties of the Gauss random process. The solid boundaries influence by step-like potential $U^\text{ext}_i$ which prohibit penetration of the $i$-th electrolyte's component below known point $z_i$:
\begin{equation}
    \label{eq:hard_wall_1}
    U^\text{ext}_i=
    \left\{
    \begin{array}{cc}
         & 0, \quad z\geq z_i \\
         & \infty, \quad z<z_i
    \end{array}
    \right.
\end{equation}
The thermodynamic contribution of the ions modeled as mixture of the hard spheres with diameters $d_i$ corresponds to Helmholtz free energy $f^{HS}$. 

\amended{
The effects of the averaged electrostatic field, the modified configuration volume and the interaction with the rough solid boundaries
can be taken into account by the Helmholtz free energy potential $F$ defined in the volume free from solid media $\int S(z) dz$ 
\begin{equation}
    \label{eq:Helmholtz_free_energy_1}
    F[\{\overline{\rho}_i\}]=\int S(z) dz \sum_i\left[U^\text{ext}_i(z)+\overline{\psi}(z) Z_i\right]\overline{\rho}_i(z)+F^{HS},
\end{equation}

$U^\text{ext}_i(z)$ is the hard boundary potential which is infinity for $z<z_i$ and zero for $z\geq z_i$, 
and $F^{HS}$ is the contribution from hard spheres interaction.
The equilibrium condition can be written in terms of c-DFT \cite{hartel2017structure} minimising the grand potential $\Omega=F-\int dz S(z)\rho(z)\mu_i$, where $\mu_i$ denotes the component chemical potential which is constant across the volume:
\begin{equation}
    \label{eq:chemical_potential_1}
    \mu_i=\frac{1}{S(z)}\frac{\delta F[\{\overline{\rho}_i\}]}{\delta \overline{\rho}_i}= U^\text{ext}_i(z)+ Z_i\overline{\psi}(z)-\mu_i^{HS},
\end{equation}
where $\mu_i^{HS}$ is the components' chemical potentials corresponding to the hard sphere repulsion. In order to calculate this term we consider an elementary layer reflecting the roughness configuration at distance $z$. More precisely, we assume the constant surface area $S(z)$ across the layer of width $\Delta z$ comparable to molecular size. This approach is similar to the Fundamental Measure Theory (FMT) \cite{roth2010fundamental}, where the hard spheres contribution is described in terms of the functions weighted over the molecular volumes.  
Therefore, inside these layers the proper volume (free from solid media) is defined as $\Delta V(z)=S(z)\Delta z$ and hard spheres mixture can be described in terms of the corresponding partition function \cite{aslyamov2019zeros}: 
\begin{equation}
    \label{eq:appendix_partition_function_1}
    Z^\text{HS}=\frac{\left(\Delta V-\sum_i v_i N_i\right)^{\sum_i N_i}}{\prod_i N_i!}
\end{equation}
where $v_i=\pi d_i^3/6$ is the molecular volume of the $i$-th component, and the number of the molecules is defined from the density distribution $N_i=\overline{\rho}_i(z) S_0\Delta z$. The hard sphere chemical potentials are defined from the partition function as 
\begin{equation}
    \label{eq:appendix_chemical_potential_2}
    \mu_i^{HS}=-k_B T\partial_{N_i}\log Z^\text{HS},
\end{equation}
}

After using of the Stieltjes formula for large $N_i$ the logarithm factor has the following form: 
\begin{equation}
    \label{eq:appendix_logarithm_1}
    \log Z^\text{HS}=\sum_i N_i \log\left(V(z)-\sum_i v_i N_i\right)-\sum_i N_i\log N_i/e.
\end{equation}
After differentiation in respect to $N_i$ expression \eqref{eq:appendix_logarithm_1} has the following form:
\begin{equation}
    \label{eq:appendix_logarithm_2}
    \partial_{N_i}\log Z^\text{HS}=\log\left(V(z)-\sum_i v_i N_i\right)-\sum_i\frac{N_i}{\left(V(z)-\sum_i v_i N_i\right)}-\log N_i\simeq-\log\frac{\overline{\rho}_i}{s(z)-\sum_iv_i\overline{\rho}_i}.
\end{equation}
Expression \eqref{eq:appendix_logarithm_2} defines the derivative of the specific energy $\partial_{\overline{\rho}_i} f^{HS}=-\partial_{N_i}\log Z^\text{HS}$. Therefore equilibrium condition \eqref{eq:chemical_potential_1} has the following form:
\begin{equation}
   \label{eq:appendix_chemical_potenital_2}
   \frac{\overline{\rho}_i}{s(z)-\sum_iv_i\overline{\rho}_i}=\theta(z-z_i)e^{\beta\mu-Z_i\beta\overline{\psi}(z)}.
\end{equation}
The chemical potential can be derived from the condition that ions are homogeneous far enough from the surface and $\overline{\psi}\to 0$:
\begin{equation}
   \label{eq:appendix_chemical_potenital_3}
   e^{\beta\mu}=\frac{\rho_0}{1-\sum_i\gamma_i},
\end{equation}
where a new parameter $\gamma_i=v_i\rho^0$ is introduced. Substituting expression expression \eqref{eq:appendix_chemical_potenital_3} one can solve the system of equations \eqref{eq:appendix_chemical_potenital_2} in respect to densities:

\begin{equation}
    \label{eq:appendix_density}
    \overline{\rho}_i(z)=\rho_0s(z)\theta(z-z_i)\frac{e^{-Z_i\beta e \overline{\psi}}}{1-\sum_i\gamma_i+\sum_i\theta(z-z_i)\gamma_i e^{-Z_i\beta e \overline{\psi}}}
    \simeq\rho_0s(z)\theta(z-z_i)\frac{e^{-Z_i\beta e \overline{\psi}}}{1-\sum_i\gamma_i+\sum_i\gamma_i e^{-Z_i\beta e \overline{\psi}}}.
\end{equation}
In this work we used the approximated expression \eqref{eq:appendix_density}  which coincides with the model \cite{kornyshev2007double} in the flat electrodes limit. Unlike combinatorial calculations used in work \cite{kornyshev2007double} for ions with the same size, we implement the partition function \eqref{eq:appendix_partition_function_1} which allows us to consider the ions with the different radii. To describe the asymmetric electrolyte, Kornyshev introduced the additional dependence $\gamma(U)$ that leads to the result, which is equivalent to the flat limit of our model \eqref{eq:appendix_density}. Therefore, the proposed approach extends the verified model \cite{kornyshev2007double} accounting for the electrolyte properties near rough electrodes.

\amended{
It is worth noting that our calculations of the hard sphere contribution results in the following specific free energy term:
\begin{equation}
    \label{eq:appendix_fhs}
    f^{HS}(z)=\sum_i \rho_i(z)\log\left[s(z)-\sum_iv_i\rho_i(z)\right].
\end{equation} 
Obtained expression \eqref{eq:appendix_fhs} can be compared with the first term of the accurate FMT approach $f^{(1)}_{FMT}(z)=\sum n_0(z)\log\left[1-n_3(z)\right]$, where $n_0$ and $n_3$ are the weighted densities \cite{roth2010fundamental}. The functions $n_0$, $n_3$ are calculated as the averages over ions volume, and become $n_0\to\sum_i\rho_i^0$, $n_1\to\sum_iv_i\rho_i^0$ at the homogeneous liquid. 
Therefore the terms \eqref{eq:appendix_fhs} coincide with $f^{(1)}_{FMT}(z)$ far enough from the wall where the fluid is homogeneous and $s(z)=1$. 
}


\amended{
\section{Notes on numerical algorithm}
In order to calculate the charge $Q$ corresponding to the potential $U$ one need to solve the equations (2), (3) supplied with proper definition of average densities distributions (6). First, substituting the density distributions (6) to the total charge density and the problem for the volume region (2) one can numerically find the volume field $\overline{\psi}_v(z)$. Second, using the obtained volume solution one can determine the field at the surface region $\overline{\psi}_s(z)$ directly from expression (3). This information is sufficient for calculation of the charge density $\overline{q}$ at both surface and volume regions using expression (6) for $\overline{\psi}_s(z)$ and $\overline{\psi}_v(z)$, respectively. The charge density, in turn, can be numerically integrated to get the total charge $Q$. In this work, the numerical solution of the nonlinear ODE (2) and integration of the charge density distribution is performed using built-in methods of Wolfram Mathematica \cite{mathematica2020}.

The total charge calculation is repeated for sufficiently fine grid of applied potentials. Next, third order smooth interpolant is constructed from tabulated dependency of $Q$ on $U$ and numerically differentiated to get the capacitance $C_d$. Interpolation and differentiation is done using built-in methods of Wolfram Mathematica \cite{mathematica2020} as well.
}

\bibliography{supercapacitors_roughness}

\end{document}